\documentclass[conference]{IEEEtran}
\IEEEoverridecommandlockouts
\usepackage{cite}
\usepackage{amsmath,amssymb,amsfonts}
\usepackage{algorithmic}
\usepackage{graphicx}
\usepackage{textcomp}
\usepackage{xcolor}
\def\BibTeX{{\rm B\kern-.05em{\sc i\kern-.025em b}\kern-.08em
    T\kern-.1667em\lower.7ex\hbox{E}\kern-.125emX}}
\begin{document}

\title{Homomorphic Encryption for Quantum Annealing with Spin Reversal Transformations}

\author{\IEEEauthorblockN{Daniel O'Malley}
\IEEEauthorblockA{\textit{EES-16} \\
\textit{Los Alamos National Laboratory}\\
Los Alamos, NM USA\\
omalled@lanl.gov}
\and
\IEEEauthorblockN{John K. Golden}
\IEEEauthorblockA{\textit{EES-16} \\
\textit{Los Alamos National Laboratory}\\
Los Alamos, NM USA \\
golden@lanl.gov}
}

\maketitle

\begin{abstract}
    Homomorphic encryption has been an area of study in classical computing for decades.
    The fundamental goal of homomorphic encryption is to enable (untrusted) Oscar to perform a computation for Alice without Oscar knowing the input to the computation or the output from the computation.
    Alice encrypts the input before sending it to Oscar, and Oscar performs the computation directly on the encrypted data, producing an encrypted result.
    Oscar then sends the encrypted result of the computation back to Alice, who can decrypt it.
    We describe an approach to homomorphic encryption for quantum annealing based on spin reversal transformations and show that it comes with little or no performance penalty.
    This is in contrast to approaches to homomorphic encryption for classical computing, which incur a significant additional computational cost.
    This implies that the performance gap between quantum annealing and classical computing is reduced when both paradigms use homomorphic encryption.
    Further, homomorphic encryption is critical for quantum annealing because quantum annealers are native to the cloud -- a third party (such as untrusted Oscar) performs the computation.
    If sensitive information, such as health-related data subject to the Health Insurance Portability and Accountability Act, is to be processed with quantum annealers, such a technique could be useful.
\end{abstract}


\section{Introduction}

Encryption is the process of encoding information so that unauthorized parties cannot interpret the information.
Historically, encryption has been used to communicate affairs of state or military matters and is now commonly used for casual activities on the internet.
The most common use of encryption is for someone, say Alice, to send data to a second party, say Bob, so that an untrusted third party, say Oscar, can deliver the data without being able to interpret the data.
Homomorphic encryption seeks to go a step further -- to allow Oscar to compute using the encoded data, producing an encoded result that Oscar cannot interpret, but that Bob and Alice can.
In the context of homomorphic encryption, Alice and Bob may be the same person.
That is, Alice sends the encoded data to Oscar, who then computes a result based upon the encoded data and sends the result to Alice.
Oscar may add value by performing computation for Alice rather than delivering the data to someone else.

It is remarkable that homomorphic encryption is even possible, and there are different levels of homomorphic encryption.
Some schemes only allow a limited set of computations to be performed, whereas fully homomorphic encryption allows for arbitrary computations to be performed.
The concept of fully homomorphic encryption was initially described decades ago \cite{rivest1978data}.
It was not until more than 30 years later that a fully homomorphic method was described \cite{gentry2009fully}, but it was very slow \cite{gentry2011implementing}.
More recent work \cite{brakerski2014leveled,gentry2013homomorphic} has brought down the computational expense significantly, but there remains significant overhead associated with these approaches.
Our approach to homomorphic encryption for quantum annealing is ``fully'' homomorphic in the sense that any computation that can be performed with a quantum annealer can also be performed with a quantum annealer using our approach.

Recently, the performance gap between quantum annealing and classical computing has been closed, but only in very narrow application spaces \cite{king2019scaling,pang2019potential}.
Homomorphic encryption for quantum annealing has the potential to close the performance gap in a broader application space.
This is because, as previously mentioned, using homomorphic encryption with classical computing comes with significant computational overhead.
However, as we demonstrate, our approach to homomorphic encryption for quantum annealing comes with little or no performance penalty.
To state this more formally, suppose that $t^{CC}_{HE}$ and $t^{QA}_{HE}$ are the times required to perform a calculation with homomorphic encryption with a classical computer and a quantum annealer, respectively.
Similarly, let $t^{CC}$ and $t^{QA}$ be the time required to perform the same calculation without homomorphic encryption with a classical computer and a quantum annealer.
Classically, $t^{CC}_{HE} \gg t^{CC}$ whereas our results show $t^{QA}_{HE} \approx t^{QA}$, so $t^{QA}_{HE} / t^{CC}_{HE} \ll t^{QA} / t^{CC}$.
Therefore, the relative performance of quantum annealing to classical computing improves when homomorphic encryption is necessary.

At present, most quantum annealing users cannot process sensitive information (such as health data, banking records, social security numbers, \emph{etc}.) without having to trust a third party.
This is because quantum annealers are primarily accessed in the cloud.
In contrast to a classical computer where private computers are abundantly available, users rarely have access to a private quantum annealer.
This makes it essential to develop methods such as the one described here if quantum annealers are to be used to process sensitive information.

The remainder of this paper is organized as follows.
Section \ref{sec:methods} describes the methods we utilize in this paper, including a brief description of quantum annealing, spin reversal transformations, and the approach to homomorphic encryption for quantum annealing that we utilize.
Section \ref{sec:results} studies the application of our approach to several diverse quantum annealing problems to demonstrate the (lack of) impact on performance.
Section \ref{sec:conclusion} describes our conclusions and gives direction for future work in this area.

\section{Methods}\label{sec:methods}
Here, we describe the methods used in this paper.
We begin with a description of the Ising model that is used by D-Wave's quantum annealers.
Next, we describe the spin reversal transformations that form the basis of our approach to homomorphic encryption for quantum annealing.
Finally, we describe our approach to homomorphic encryption for quantum annealing.

\subsection{Quantum Annealing}\label{sec:qa}
Quantum annealing is a heuristic optimization algorithm, similar to simulated annealing \cite{kirkpatrick1983optimization}, that seeks to find optimal solutions faster than classical methods by exploiting quantum fluctuations \cite{kadowaki1998quantum}.
The adiabatic theorem provides theoretical guarantees of convergence to the optimal state under certain conditions (notably, slow evolution of the Hamiltonian) \cite{morita2008mathematical}.
In practice, these assumptions are generally violated.
For example, with the D-Wave quantum annealers \cite{johnson2011quantum} that we use here, the anneal process is often performed quickly -- violating the assumptions of the adiabatic theorem.

\begin{figure}[htbp]
    \centerline{\includegraphics[width=0.5\textwidth]{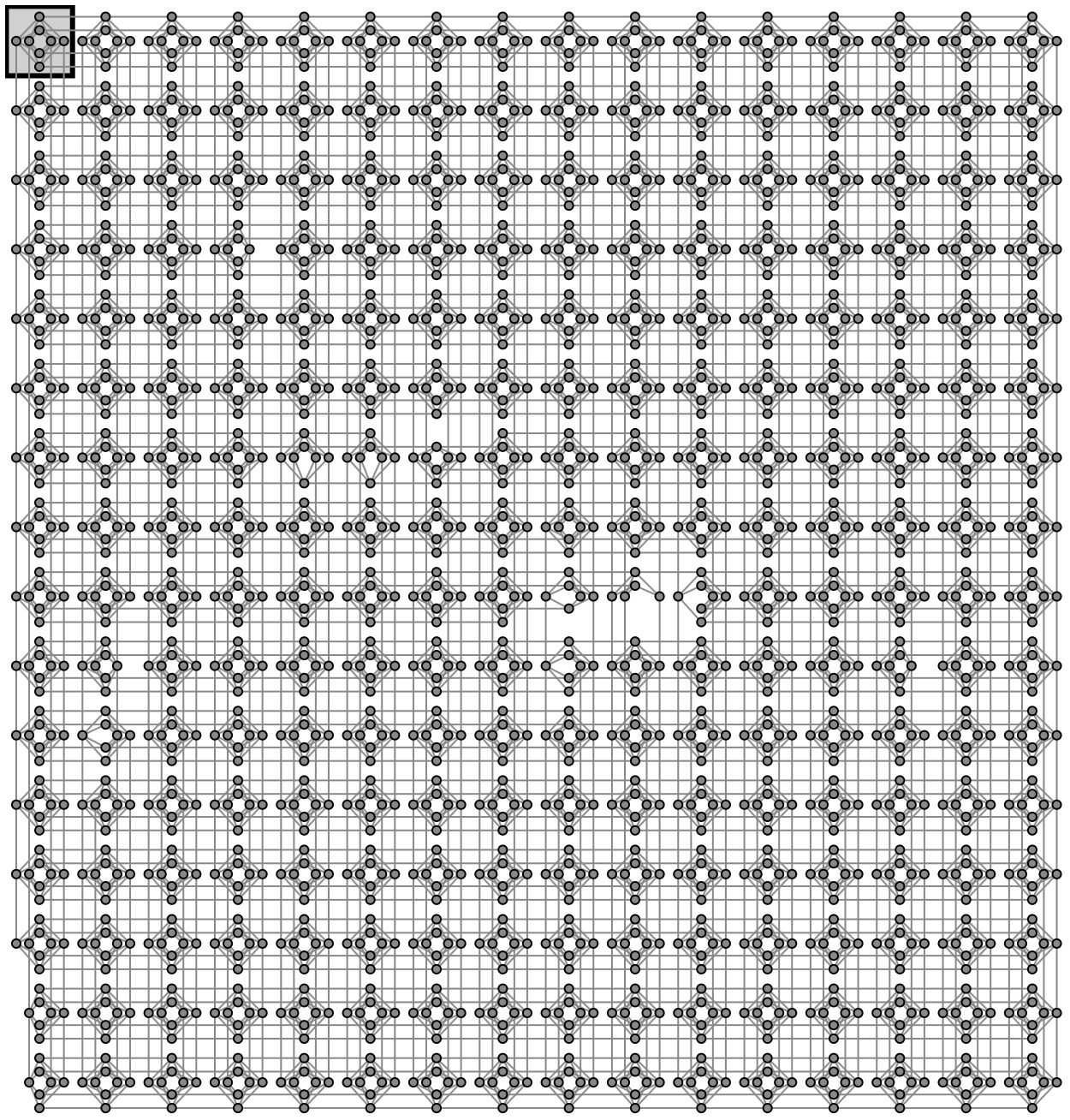}}
\caption{The Chimera subgraph associated with the D-Wave 2000Q hardware at Los Alamos National Laboratory.
    Each vertex in the graph corresponds to a qubit, and each edge corresponds to a coupler between two qubits.
    This graph determines the sparsity of the matrix $J_{ij}$.
    When there is a coupler between qubits $i$ and $j$, $J_{ij}$ is permitted to be nonzero, but if there is no coupler, then $J_{ij}=0$.
    Each qubit is coupled to at most six other qubits.}
    \label{fig:hardwaregraph}
\end{figure}

The basic input to a D-Wave quantum annealer (sometimes referred to as a ``quantum machine instruction'') is a vector $\mathbf{h}=(h_i)$ and a matrix $J=(J_{ij})$.
The matrix, $J$, is sparse with a sparsity pattern defined by the connectivity graph associated with the annealer's qubits.
For existing D-Wave quantum annealers, this is based on a Chimera graph \cite{boothby2016fast} (see Fig. \ref{fig:hardwaregraph}).
The vector, $\mathbf{h}$ and matrix $J$ define an Ising Hamiltonian
\begin{equation}
    H_p = \sum_{i=1}^{n} h_i \sigma_i^z + \sum_{i,j=1}^{n} J_{ij} \sigma_i^z \sigma_j^z
    \label{eq:problemhamiltonian}
\end{equation}
The D-Wave quantum annealer evolves the Hamiltonian over time
\begin{equation}
    H(t) = \Gamma(t) \sum_{i=1}^n \Delta_i \sigma_i^x + \Lambda(t) H_p
\end{equation}
with $\Gamma(t)$ decreasing to zero in time and $\Lambda(t)$ increasing from zero in time.
From a practical perspective, the quantum annealer can be thought of as minimizing a function of the form
\begin{equation}
    f(\mathbf{s}) = \sum_{i=1}^{n} h_i s_i + \sum_{i,j=1}^{n} J_{ij} s_i s_j
    \label{eq:isingfunction}
\end{equation}
where each spin, $s_i$, is either $+1$ of $-1$.
A more accurate description of the behavior of the annealer is that it is drawing from a distribution that preferentially samples values of $\mathbf{s}$ that make $f(\mathbf{s})$ small.
This distribution can often be well-approximated by a Boltzmann distribution where Eq. \ref{eq:isingfunction} defines the energy.

\subsection{Spin Reversal Transformations}
We now describe the process of performing a spin reversal transformation (sometimes called a gauge transformation).
Conceptually, the spin reversal transformation modifies the Ising Hamiltonian (Eq. \ref{eq:problemhamiltonian}) so that samples obtained with the modified Hamiltonian can be mapped back to samples from the original Hamiltonian.
Further, the energy will be the same for both samples with respect to each of their Hamiltonians.
This energy equivalence provides some theoretical confidence that, at least in the sense of the Boltzmann approximation mentioned previously, the performance of the quantum annealer is not degraded by the spin reversal transformation.

The spin reversal transformation works as follows.
Let $\mathbf{x}=(x_i)$ be a binary string, and define
\begin{equation}
    H_p^* = \sum_{i=1}^{n} (-1)^{x_i} h_i \sigma_i^z + \sum_{i,j=1}^{n} (-1)^{x_i} (-1)^{x_j} J_{ij} \sigma_i^z \sigma_j^z
\end{equation}
and a corresponding energy
\begin{equation}
    f^*(\mathbf{s}) = \sum_{i=1}^{n} (-1)^{x_i} h_i s_i + \sum_{i,j=1}^{n} (-1)^{x_i} (-1)^{x_j} J_{ij} s_i s_j
\end{equation}
Given a sample, $\mathbf{s}^*=(s_i^*)$, we define a corresponding sample, $\mathbf{s}=(s_i)$, where
\begin{equation}
    s_i = (-1)^{x_i} s_i^*
    \label{eq:sgauge}
\end{equation}
With this notation, our previous statement about the energy equivalence can be stated more precisely,
\begin{eqnarray}
    f^*(\mathbf{s}^*) &=& \sum_{i=1}^{n} (-1)^{x_i} h_i s_i^* + \sum_{i,j=1}^{n} (-1)^{x_i} (-1)^{x_j} J_{ij} s_i^* s_j^* \nonumber\\
             &=& \sum_{i=1}^{n} (-1)^{x_i} h_i (-1)^{x_i}s_i \nonumber\\
             && + \sum_{i,j=1}^{n} (-1)^{x_i} (-1)^{x_j} J_{ij} (-1)^{x_i} s_i (-1)^{x_j} s_j \nonumber\\
             &=& \sum_{i=1}^{n} h_i s_i + \sum_{i,j=1}^{n} J_{ij} s_i s_j \nonumber\\
             &=& f(\mathbf{s}) \label{eq:fequivalance}
\end{eqnarray}
We will also use the notation
\begin{equation}
    h_{i}^* = (-1)^{x_i} h_i
    \label{eq:hgauge}
\end{equation}
and
\begin{equation}
    J_{ij}^* = (-1)^{x_i}(-1)^{x_j} J_{ij}
    \label{eq:jgauge}
\end{equation}
This transformation between $(H_p^*, f^*, J^*, h^*, \mathbf{s}^*)$ and $(H_p, f,  J, h,\mathbf{s})$ is called a spin reversal transformation or gauge transformation.

Spin reversal transformations have been proposed as a means to improve the performance of the quantum annealer \cite{pelofske2019optimizing,dwave2020usage}.
These performance improvements are based on the practicalities of the quantum annealer.
They seek to mitigate biases in the realization of the Hamiltonian specified by the user.
By applying numerous spin reversal transformations, the biases in the Hamiltonian specification can be averaged out in some sense.
Some spin reversal transformations may degrade performance slightly while others enhance performance slightly.
The variability in performance should decrease as the integrated control errors associated with the hardware are reduced.
We do not require the spin reversal transformation to provide a performance enhancement -- merely maintaining approximately the same performance is what we aim to demonstrate.
However, methods for improving performance using spin reversal transformations can be used on top of our approach to homomorphic encryption for quantum annealing.
The benefits of these approaches would still apply and would not hinder our approach.

\subsection{Homomorphic Encryption for Quantum Annealing}
\begin{figure}[htbp]
    \centerline{\includegraphics[width=0.5\textwidth]{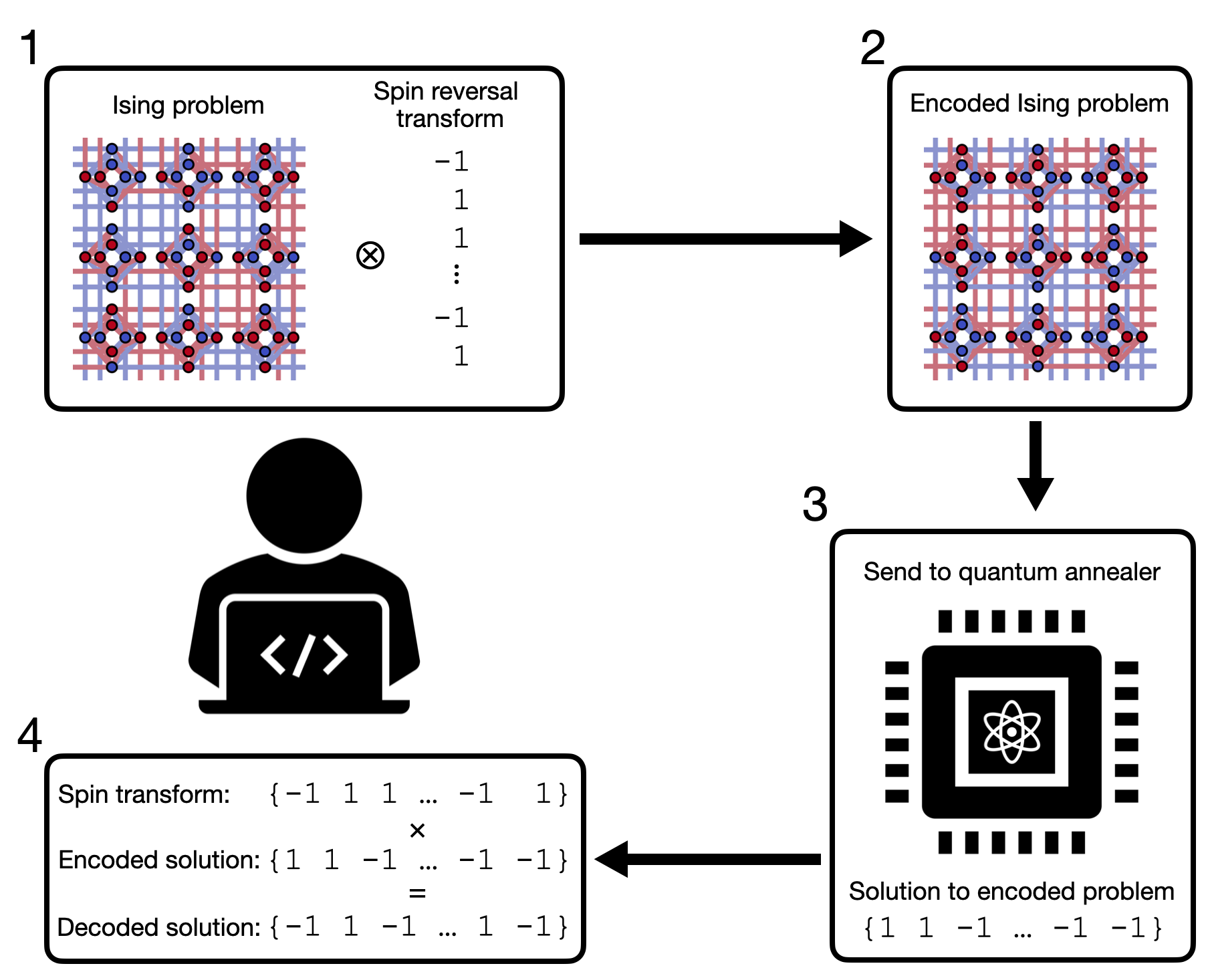}}
\caption{The workflow for our approach to homomorphic encryption for quantum annealing.
    (1) The user constructs an Ising problem, generates a secret key, and applies a spin reversal transformation the Ising problem based on the key to obtain an encoded Ising problem (2).
    Then the encoded Ising problem is sent to and solved by the quantum annealer (3).
    Finally, the solution is retrieved from the quantum annealer and decoded by applying the spin reversal transformation to the solution (4).}
    \label{fig:workflow}
\end{figure}

Our approach to homomorphic encryption utilizes the spin reversal transformation as the key mechanism.
The basic approach is as follows (see Fig. \ref{fig:workflow}).
\begin{enumerate}
    \item The user constucts an Ising model, $(h_i)_{i=1}^n$ and $(J_{ij})_{i,j=1}^n$ (where $n$ is the number of qubits), for a problem of interest.
    \item The user generates a random string of bits, $\mathbf{x}=(x_i)_{i=1}^n$ that acts as a secret key.
    \item The user applies the spin reversal transformation to obtain $(h_i^*)_{i=1}^n$ and $(J_{ij}^*)_{i,j=1}^n$ using the definitions in Eqs. \ref{eq:hgauge} and \ref{eq:jgauge}.
    \item The user sends $(h_i^*)_{i=1}^n$ and $(J_{ij}^*)_{i,j=1}^n$ to a third party that controls the quantum annealer.
    \item The third party obtains $m$ samples from the quantum annealer. We denote the $j^{\mathrm{th}}$ sample, $\mathbf{s}_j^*$ and the $i^{\mathrm{th}}$ component of $\mathbf{s}_j^*$ as $s_{ij}^*$.
    \item The user retrieves the samples, $\mathbf{s}_j^*$, from the third party.
    \item The user applies the spin reversal transformation to each sample, $\mathbf{s}_j^*$, to obtain $\mathbf{s}_j$ using the definition in Eq. \ref{eq:sgauge}.
\end{enumerate}
At the end of this process, the user will have found a sample that minimizes $f$ if and only if the annealer obtained a sample that minimizes $f^*$.
Similarly, the distribution of energies of the samples obtained is unchanged by the spin reversal transformation.
These assertions are both a straightforward consequence of the fact that $f^*(\mathbf{s}^*)=f(\mathbf{s})$ (equation \ref{eq:fequivalance}).

The sample obtained from the quantum annealer, $\mathbf{s}^*$ cannot be transformed into $\mathbf{s}$ without knowledge of the secret key, $\mathbf{x}$.
However, this secret key is not a one-time pad, because it is also used to transform the Ising problem ($\mathbf{h}$ and $J$).
We leave the analysis of the Ising problem to infer the secret key, $\mathbf{x}$, for future work.

Many problems that are solved with a quantum annealer use multiple physical qubits to represent a single logical qubit through an embedding process \cite{boothby2016fast}.
This is an approach to deal with the limited connectivity associated with the graph in Fig. \ref{fig:hardwaregraph}.
While further analysis may be required and the most secure behavior may have some problem dependence, we recommend performing the gauge transformation on the logical Ising problem rather than the physical Ising model.
For embedded problems, this would mean inserting an extra step between steps 3 and 4 above where the Ising model is embedded and a corresponding step between steps 6 and 7 where the samples are unembedded.
In cases where multiple rounds of annealing are used for the same problem, we recommend using the same spin reversal transformation repeatedly in this case.

\section{Results}\label{sec:results}
In this section, we detail quantitative results from applying homomorphic encryption on a variety of established Ising problem classes amenable to quantum annealing. 
Our methodology for testing is to generate an Ising problem from each class and evaluate it on the D-Wave 2000Q hardware at Los Alamos National Laboratory. 
We begin by generating $10^4$ samples of the problem with no spin reversal transformation applied. 
From these samples, we plot the cumulative distribution functions (CDFs).
The CDF is the function that maps a value, $x$, to the probability that a random sample has energy less than or equal to $x$.
We then generate and apply ten different spin reversal transforms and evaluate their CDFs (again using $10^4$ samples).
We aim to show that the CDFs from the untransformed case approximately match those associated with the different spin reversal transformations.
It is not expected that there be an exact match, as sampling from the quantum annealer is inherently stochastic, and the integrated control errors impact different spin reversal transformations differently.
However, it is expected that there should be reasonably close agreement between the CDFs for the transformed and untransformed problems.

We explore three Ising problems taken from problems related to nonnegative/binary matrix factorization \cite{omalley2018nbmf}, hydrologic inverse analysis \cite{omalley2018hydro}, and RAN1 benchmarks \cite{jack2016sampling}.
These problems were chosen for their diverse characteristics.
The nonnegative/binary matrix factorization problem utilizes an embedding of a complete graph, which requires many physical qubits to represent a single logical qubit.
The hydrologic inverse problem depends upon a structured embedding that utilizes a small number (at most 4) of physical qubits per logical qubit.
The RAN1 problem does not utilize an embedding and uses one physical qubit per logical qubit.
There is also diversity in the dynamic range properties with the hydrologic inverse problem utilizing a wide dynamic range, the RAN1 problem a dynamic range that is perfectly matched to the hardware, and the nonnegative/binary matrix factorization problem being somewhere in between.

\subsection{Nonnegative/Binary Matrix Factorization Problem}
\begin{figure}[htbp]
    \centerline{\includegraphics[width=0.5\textwidth]{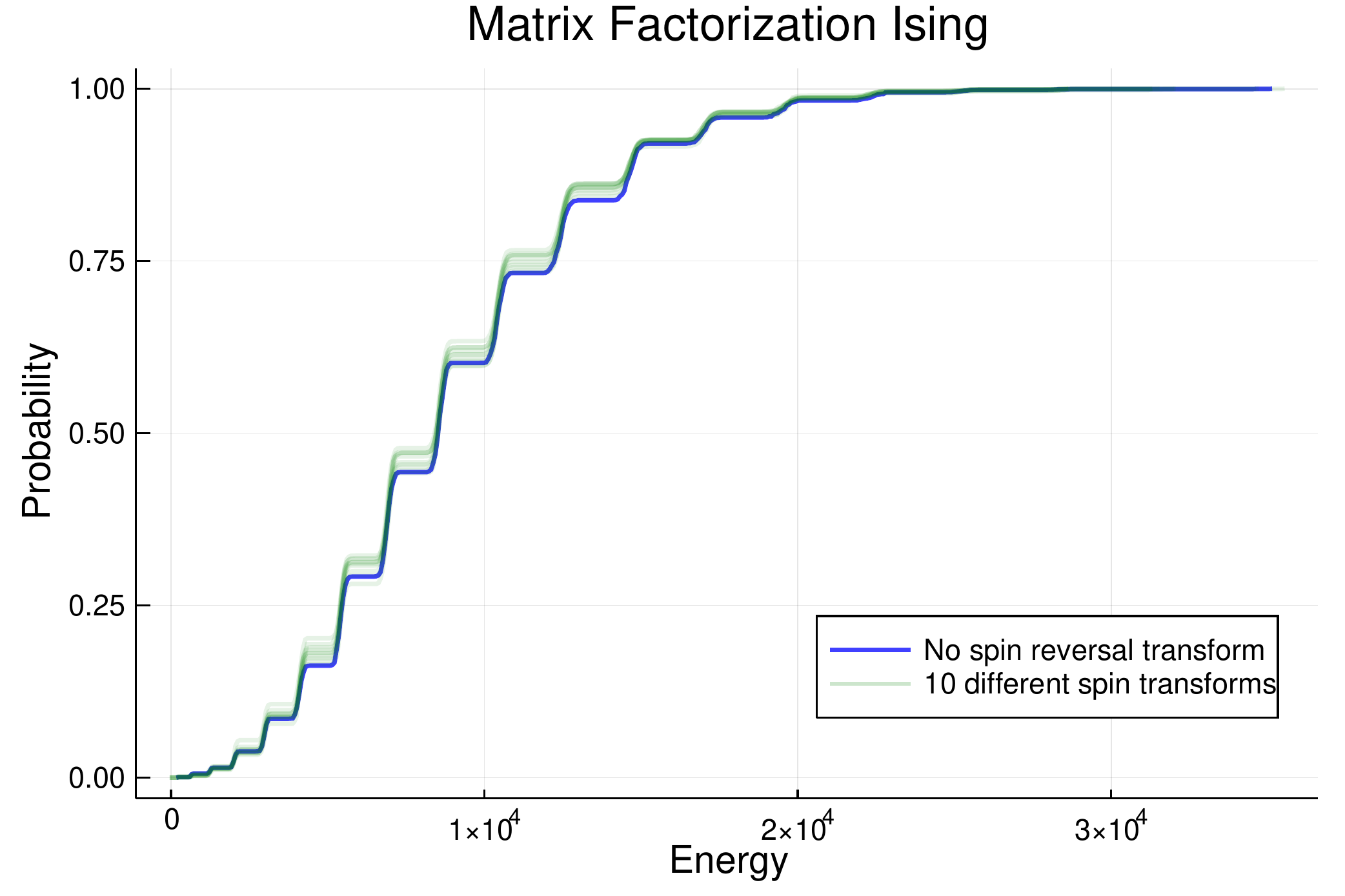}}
    \caption{The cumulative density function of the energies from an Ising model that arises in a nonnegative/binary matrix factorization problem is shown.
    The similarity between the green curves (which use a spin reversal transformation) and the blue curve indicates that the performance is unhindered by the spin reversal transformation.}
    \label{fig:cdf_faces}
\end{figure}

The nonnegative/binary matrix factorization problem was introduced in \cite{omalley2018nbmf}.
The goal of this problem is to take a real-valued $n \times m$ matrix $A$ and find $B$ and $C$ such that 
\begin{equation}
A \approx B C
\end{equation}
where $B$ is a nonnegative $n \times k$ matrix and $C$ is a binary $k \times m$ matrix.
After randomly initiating a seed matrix $C^{(0)}$, each iteration follows an alternating least squares approach:
\begin{align}
& \text{find } B^{(i)} = \text{arg min} \| A - X C^{(i-1)}\|, \\
& \text{find } C^{(i)} = \text{arg min} \| A - B^{(i)} X\|, \label{eq:binary-matrix-opt}
\end{align}
where eq.~(\ref{eq:binary-matrix-opt}) can be efficiently solved on a quantum annealer.
For our study, we randomly chose an Ising problem generated during the evaluation of the nonnegative/binary matrix factorization algorithm applied to a matrix composed of data from facial images \cite{cbcl2017cbcl}.
This Ising problem features 35 logical variables, which are embedded on 437 physical qubits with 1082 couplers.
As shown in Fig.~\ref{fig:cdf_faces}, the resulting Ising problems feature a moderately discretized energy spectrum.
The average difference in probability across the entire CDF between the untransformed and transformed results for this problem was $-0.6\%$.
The minus sign indicates that the transformed problems slightly outperformed the original problem, i.e., the chance that an untransformed sample is below a given energy is slightly less than the chance that a transformed sample is below that energy.  

\subsection{Hydrology Inverse Problem}
\begin{figure}[htbp]
    \centerline{\includegraphics[width=0.5\textwidth]{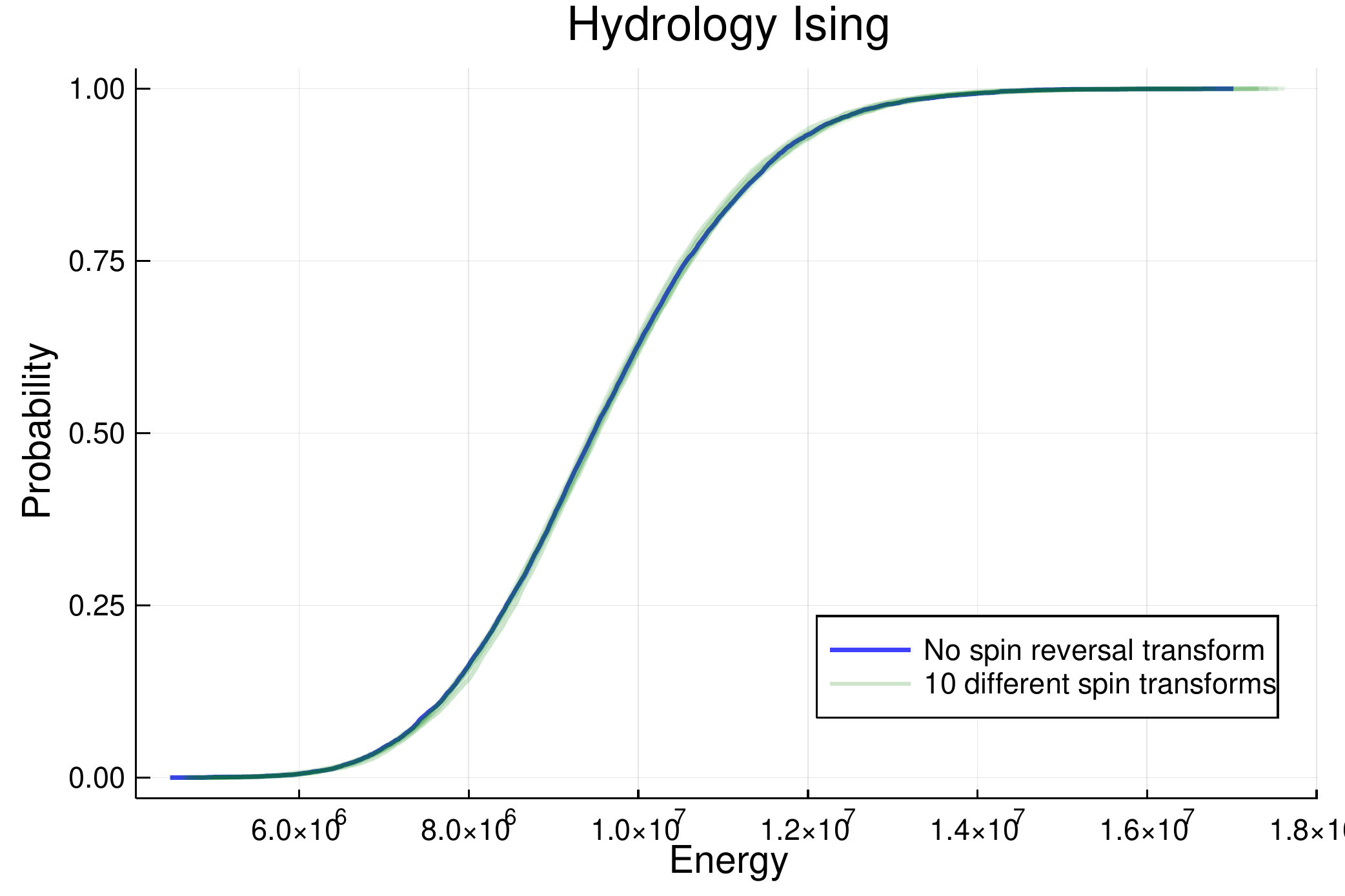}}
\caption{The cumulative density function of the energies from an Ising model that arises in a hydrologic inverse problem is shown.}
    \label{fig:cdf_hydro}
\end{figure}

Next, we study a class of Ising problems useful in solving a class of hydrologic inverse problems, first introduced in a quantum annealing context in \cite{omalley2018hydro}.
An example application of this problem is to determine the constituent materials of an aquifer given pressure readings from wells spread across the aquifer.
In this case we are solving the partial differential equation (PDE)
\begin{equation}\label{eq:pressure}
    \nabla \cdot (\mathbf{k} \cdot \nabla h) = 0.
\end{equation}
Here $\mathbf{k}$ is a vector describing the permeability at each location in an aquifer, and $h$ is pressure of the water in the aquifer.
The exact formulation of this PDE in terms of an Ising problem is detailed in \cite{omalley2018hydro}.
This Ising problem is characterized by a very smooth energy spectrum, seen in Fig.~\ref{fig:cdf_hydro}. 
To study this problem, we generated a random aquifer profile (i.e., a random $\mathbf{k}$), computed the resulting pressure readings (i.e., $h$), and then used the values of $h$ to reconstruct the aquifer permeabilities (i.e., $\mathbf{k}$) via quantum annealing. 
In this case, the Ising problem features 544 logical variables, embedded via 2048 physical qubits and 6016 couplers (using the D-Wave's Virtual Full Yield Chimera solver).
The average difference in probability across the entire CDF between the untransformed and transformed results for this problem was $-0.02 \%$
Again we see that the transformed problem slightly outperformed the original. 

\subsection{RAN1 Problem}
\begin{figure}[htbp]
    \centerline{\includegraphics[width=0.5\textwidth]{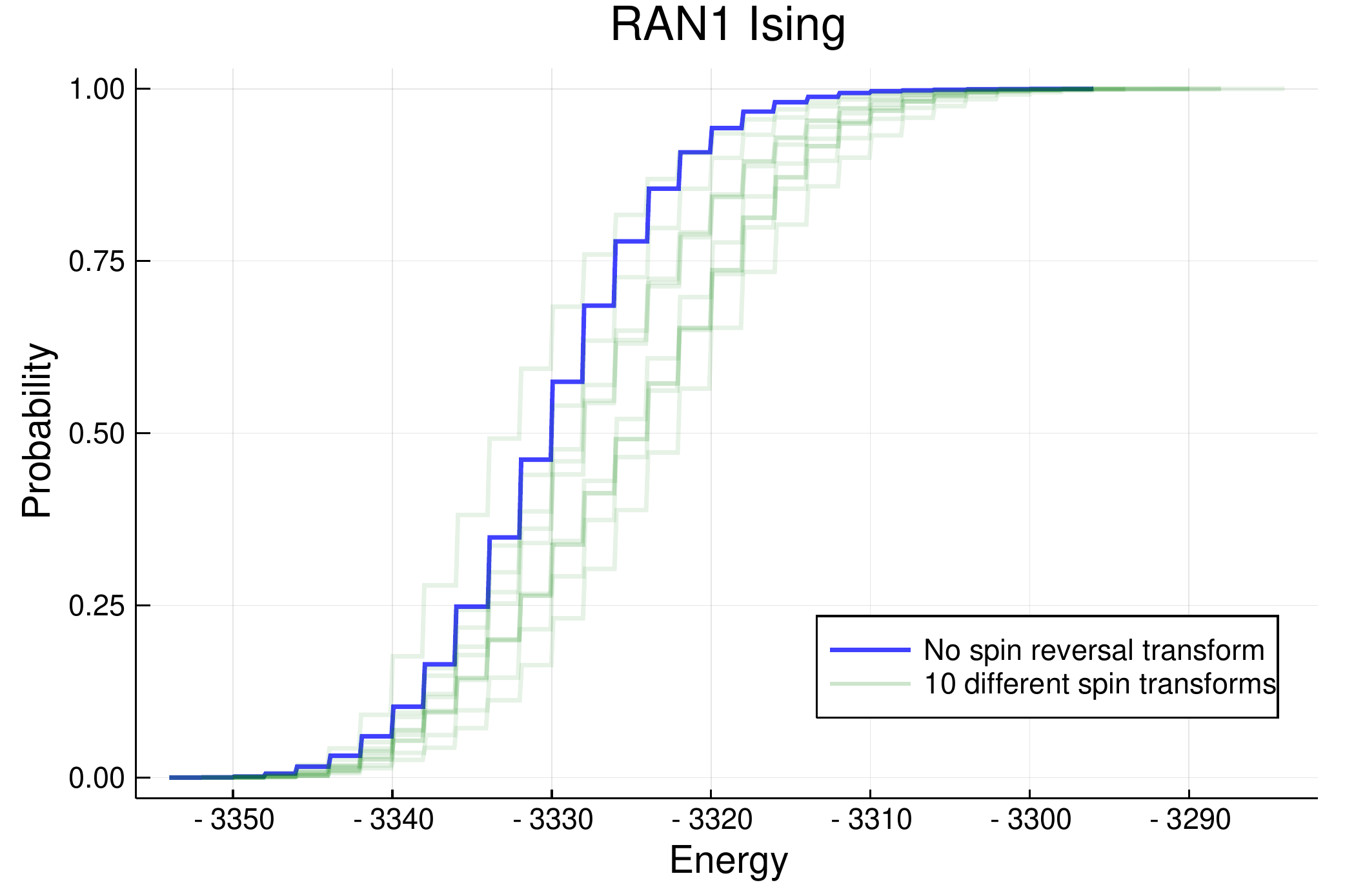}}
\caption{The cumulative density function of the energies from an Ising model that arises in a RAN1 problem is shown.
    There is more apparently variability in the performance here, but this appears to be due to statistical fluctuations.}
    \label{fig:cdf_ran1}
\end{figure}

Our final class of problems is known as RAN1, which has $h_i = 0$ and $J_{i,j}$ randomly drawn from $\{-1,1\}$.
These problems are commonly used as a benchmark in the quantum annealing community \cite{jack2016sampling}.
We studied several instantiations with 5828 active couplers. 
These problems exhibit several interesting features. 
First is the highly discretized energy spectrum, as visible in Fig.~\ref{fig:cdf_ran1}. 
Second is the high variance depending on the spin reversal transform used. 
The third is the relatively compact range of energies sampled by the D-Wave.
Compared to Figs.~\ref{fig:cdf_faces} and \ref{fig:cdf_hydro}, the energies of the RAN1 problem are all quite close to optimal.
This compact energy spectrum enhances the apparent variability. 
Still, the average difference in probability across the entire CDF between the untransformed and transformed results for this problem was $6.1\%$.
Here, the positive number indicates that the untransformed problem slightly outperformed the transformed problem.

While the transformed versions on average performed slightly worse than the original version, we point out that this appears to simply be due to statistical variability.
This is based on experience from performing the analysis repeatedly as well as the following argument.
Suppose $\mathbf{h}$ and $J$ form a RAN1 problem and $\mathbf{h}^*$ and $J^*$ is the spin reversal transformed version of the problem based on the key $\mathbf{x}$.
The transformed problem, $\mathbf{h}^*$ and $J^*$, is also a realization of a RAN1 problem that is equally likely to the untransformed version.
Now suppose that $\mathbf{h}^*$ and $J^*$ had been sampled as the RAN1 problem instead of $\mathbf{h}$ and $J$.
In this case, $\mathbf{h}$ and $J$ would be the transformed problem (based on the same key $\mathbf{x}$).
Any performance that is gained or lost in the first case (where $\mathbf{h}$ and $J$ is the sampled RAN1 problem) is lost or gained, respectively, in the second case (where $\mathbf{h}^*$ and $J^*$ is the sampled RAN1 problem).
Therefore, there is no performance gained or lost in a statistical sense over the class of RAN1 problems.

\section{Conclusion \& Future Work}\label{sec:conclusion}
We have described an approach to homomorphic encryption that applies to quantum annealing.
The approach exploits the theoretical invariance of the quantum annealing process with respect to spin reversal transformations.
We also demonstrated that these spin reversal transformations do not significantly degrade the quantum annealer's performance by exploring three example applications.
For some of these spin reversal transformations, the performance is improved slightly, and for others, it is degraded slightly.
On average, the performance when using a random spin reversal transformation is about the same as when no spin reversal transformation is used.
This can help quantum annealing close the performance gap with classical computing in applications where homomorphic encryption is required.

In this process, a secret key is generated and never shared.
The samples that are obtained by the quantum annealer (and can be observed by the third party) are encrypted with the secret key.
The same secret key is also used to transform the Ising model that is sent to the quantum annealer (which can also be observed by the third party).
In future work, it is important to identify classes of problems for which the Ising model does not betray information about the secret key.
One such class may be the RAN1 class because the spin reversal transformation simply transforms one random RAN1 problem into another RAN1 problem.
However, further work would be required to make this hypothesis rigorous, and to identify other classes that have more computational significance.

The results and description of the methods here are focused on forward quantum annealing, where the annealing process starts in the uniform superposition of all states.
There is another quantum annealing technique called reverse quantum annealing.
In reverse quantum annealing, a classical state, $\mathbf{s}_0$, is sent to the quantum annealer in addition to $\mathbf{h}$ and $J$.
The reverse annealing process starts in the classical state, $\mathbf{s}_0$ rather than the uniform superposition.
Reverse quantum annealing is more of a local search heuristic than a global search heuristic \cite{chancellor2017modernizing} and provides a mechanism to refine a solution.
It was introduced in D-Wave quantum annealers as a second ``quantum machine instruction'' with the first instruction described in Section \ref{sec:qa}.
Our approach can be readily adapted to reverse quantum annealing by merely applying the spin reversal transformation to $\mathbf{s}_0$ before sending that to the annealer -- the rest of the process proceeds as described in Section \ref{sec:methods}.

Another possible extension would be to a scenario where two users, say Alice and Bob, have Ising models that need to be combined.
This would be important, e.g., in a scenario where Alice and Bob each have different sets of information about a patient and the combined information is needed, e.g., to make a diagnosis.
With quantum annealing, this can be performed by having Oscar sum the Ising models provided separately by Alice and Bob.
In this scenario, however, Oscar gets to examine both of the Ising models, which would be undesirable if the information were sensitive.
Another option is for Alice and Bob to establish a shared private key and both encrypt their Ising models using the same key.
Alice and Bob separately encode their Ising models using the shared key.
Oscar can then sum the encoded Ising models and perform the computation without being able to directly observe the original Ising models.
In this scheme, Oscar would only have to be trusted not to share Alice's Ising model with Bob or Bob's Ising model with Alice.

We have described an approach to homomorphic encryption in the context of quantum annealing.
There are also other computational architectures \cite{mcmahon2016fully,bohm2019poor,aramon2019physics} that exploit an Ising model of computation.
The method can be readily applied to those architectures as well.
However, testing would be required for each of these other architectures to evaluate their performance when using a spin reversal transformation.

\section*{Acknowledgment}

We wish to acknowledge Steve W. Poole for his insights and discussions, which greatly assisted us during the preparation of this manuscript.
We also thank the anonymous reviewers, whose insightful comments significantly improved the manuscript.


\bibliographystyle{IEEEtran}

\end{document}